\documentclass[aps,twocolumn]{revtex4}
\usepackage{epsfig}

\newcommand{\etal}{{\it{et al.}}$\ $}

\begin{document}

\title{Theory of Feshbach molecule formation in a dilute gas during a magnetic field ramp}

\author{J. E. Williams, N. Nygaard\cite{danish}, C. W. Clark}

\affiliation{Electron and Optical Physics Division, National Institute of Standards and Technology, Gaithersburg, Maryland 20899-8410}

\begin{abstract}
Starting with coupled atom-molecule Boltzmann equations, we develop a simplified model to understand molecule formation observed in recent experiments. Our theory predicts several key features: (1) the effective adiabatic rate constant is proportional to density; (2) in an adiabatic ramp, the dependence of molecular fraction on magnetic field resembles an error function whose width and centroid are related to the temperature; (3) the molecular production efficiency is a universal function of the initial phase space density, the specific form of which we derive for a classical gas.  Our predictions show qualitative agreement with the data from [Hodby \etal, Phys. Rev. Lett. {\bf{94}}, 120402 (2005)] without the use of adjustable parameters.
\end{abstract}

\maketitle

By ramping a magnetic field across a Feshbach resonance, loosely bound diatomic molecules have been created in two-component Fermi gases above~\cite{Regal2003a,Strecker2003a} and below~\cite{Greiner2003a,Regal2004a,Bourdel2004a} the superfluid transition temperature. In this process, there appear to be two distinct relaxation timescales: a ``two-body" timescale in the range of $10$'s to $100$'s of microseconds during which molecules form, and a longer ``many-body" timescale in the range of $10$'s of milliseconds during which a condensate forms~\cite{Regal2004a,Zwierlein2005a}. A critical unresolved riddle emerging from these experiments is that the short-time molecular formation physics can not be fully accounted for by theories that treat the Feshbach ramp as a purely two-body process~\cite{Mies2000a,Pazy2004a}, suggesting a many-body treatment is warranted~\cite{Javanainen2004a,Chwedenczuk2004a,Williams2004a}. In this letter, we argue that in the above-cited experiments, multiple collisions per atom occur during the ramp. Thus, Feshbach molecule formation in a thermal gas must be viewed as a collisional relaxation process, which we describe using kinetic theory~\cite{Williams2004a,Nygaard2005b} and thermodynamics~\cite{Kokkelmans2004a,Carr2004a,Williams2004b}.

% You need to introduce the notion of "saturation", that is, a maximum of production efficiency of less than one. This got lost somewhere along the way.
Several key experimental quantities in need of a comprehensive theoretical understanding are illustrated qualitatively in Fig.~1. As the magnetic field is tuned across the resonance, the molecular fraction $\chi\equiv 2N_M/N_{\rm{tot}}$ increases to a maximum value $\chi_{\dot B}$, where $N_{\rm{tot}}$ and $N_M$ are the total and molecular populations. This behavior is represented in Fig.~1a by the error function $\chi_{\rm{erf}}(B) = \chi_{\dot B}\{1 - {\rm{erf}}[\sqrt 2(B- B_{\rm{cen}})/\delta B] \}/2$, which has been used to fit experimental data~\cite{Regal2003a,Regal2004a}. Fig.~1b shows how the production of molecules increases and then saturates to a value less than unity as the ramp rate is decreased. An exponential function $\chi_{\dot B} = \chi_0[1 - \exp(-\alpha/|\dot B|)]$ gives a reasonable fit to data~\cite{Regal2003a,Strecker2003a,Regal2004a}, however, a power-law form can not be ruled out~\cite{Pazy2005a}. The rate constant $\alpha$ determines whether the sweep is adiabatic and $\chi_0$ is the maximum production efficiency obtained in the adiabatic limit $\alpha/|\dot B| \gg 1$. In this letter, we
identify the root cause of the saturation effect and determine the scaling behavior of the measured properties $B_{\rm{cen}}$, $\delta B$, $\alpha$ and $\chi_0$.

\begin{figure}[t]
\begin{center}
\includegraphics[scale=0.39]{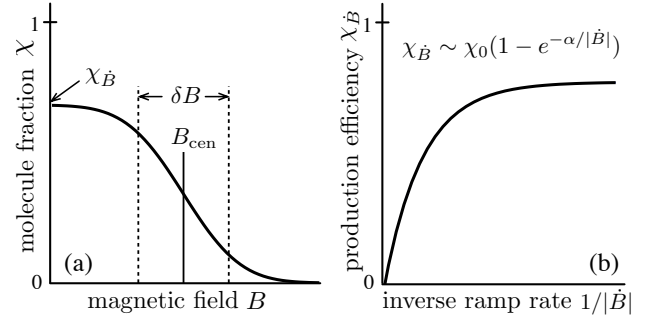}
\caption{Illustrations of the molecular fraction versus (a) magnetic field $B$ and (b) sweep rate $|\dot B|$. }
\end{center}
\end{figure}

A useful idealization that is commonly invoked to understand Feshbach molecule formation is to scale the system size down to two isolated trapped atoms initially in their ground state. This description is amenable to a straight-forward application of the Landau-Zener (LZ) level crossing theory~\cite{Mies2000a}, for which the effect of the magnetic field ramp is easy to conceptualize: the state of the atom pair tracks the lowest energy state during the ramp, eventually crossing threshold to become a true two-body bound state. In the adiabatic limit, the LZ model predicts that at the end of the sweep the probability of finding the atom pair in a bound molecular state is unity, corresponding to a production efficiency $\chi_0 = 1$. For finite temperature gases, this two-atom LZ model is not directly applicable, and thus one does not necessarily expect that a unit conversion efficiency can be achieved. 

Nevertheless, is there a way to incorporate the two-atom LZ theory into a model that takes into account the presence of many atoms in a trapped gas? The key to answering this is to realize that in a dilute gas there is an upper bound on the ramp time $t_{\rm{ramp}}$ set by the average time between collisions $\tau_{\rm{col}}$: if $t_{\rm{ramp}} \gg \tau_{\rm{col}}$, an individual atom may undergo multiple collisions during the ramp, and this physics is not taken into account by the LZ approach. In the opposite limit $t_{\rm{ramp}} \ll \tau_{\rm{col}}$, each atom will likely encounter at most a single collision partner -- its nearest neighbor, so that the LZ theory can be applied to isolated pairs of atoms distributed throughout the gas. Models working in this type of ``static gas" approximation (SGA) predict a maximum production efficiency of $\chi_0 = 0.5$~\cite{Pazy2004a,Chwedenczuk2004a}.  Although this efficiency limit is consistent with the first two Feshbach sweep experiments in a Fermi gas~\cite{Regal2003a,Strecker2003a}, the more extensive recent study of Hodby \etal~\cite{Hodby2005a} measured values of $\chi_0$ in the range $0.1 - 0.9$, in disagreement with the SGA prediction.

We can understand this apparent breakdown of the SGA by estimating $\tau_{\rm{col}}$ for typical experimental parameters using $\tau_{\rm{col}}^{-1}=n\sigma \bar v_{\rm{rel}}$, where $\sigma$ is the collisional cross section and $\bar v_{\rm{rel}}$ is the average relative velocity. In the unitarity limit, $\sigma = 4\pi a^2/(1 + k^2 a^2)$, which can be approximated by $\sigma = 4 \pi/k_{\rm{F}}^2$ close to the resonance where the scattering length $a$ diverges. Here $k_{\rm{F}}$ is the Fermi wave vector. Estimating the density as $n \approx (2m k_{\rm{B}} T_{\rm{F}}/\hbar^2)^{3/2}/6\pi^2$, where $m$ is the atomic mass, and taking $\bar v_{\rm{rel}}=4\sqrt{k_{\rm{B}}T_{\rm{F}}/\pi m}$, the collisional time becomes $\tau_{\rm{col}} \approx \hbar/k_{\rm{B}} T_{\rm{F}}$~\cite{Zwierlein2005a}. In the JILA experiments~\cite{Hodby2005a}, typical Fermi temperatures are $T_{\rm{F}}=350$ nK, giving $\tau_{\rm{col}}=20 \ \mu$s. With $t_{\rm{ramp}}\gtrsim 50 \ \mu$s in these experiments, multiple collisions per atom occur during the ramp and an alternative to the SGA is necessary.

To go beyond the SGA, Hodby \etal devised a stochastic phase space sampling (SPSS) model based on the assumption that the probability of two atoms forming a molecule depends on their proximity in phase space~\cite{Hodby2005a}. With a single fitting parameter -- a cut-off radius in phase space -- the phenomenological SPSS model predicts that that $\chi_0$ is a universal function of $\rho_{0i}$ that agrees remarkably well with their experimental data. In this letter, we provide a theoretical backdrop for understanding this behavior and derive from first principles the universal functional form for $\chi_0(\rho_{0i})$.

Rather than try to adapt the two-atom LZ theory to the case of a finite-temperature gas, we have developed an entirely different many-body approach that is appropriate for the regime $t_{\rm{ramp}} \gg \tau_{\rm{col}}$. Using the Keldysh nonequilibrium Green's function formalism~\cite{Keldysh1964a}, we derived a pair of coupled Boltzmann equations describing the dynamical evolution of the atomic $f_A({\mathbf{p}},{\mathbf{r}},t)$ and molecular $f_M({\mathbf{p}},{\mathbf{r}},t)$ distribution functions~\cite{Nygaard2005b}; we assume an equal spin mixture $f_A \equiv f_\uparrow=f_\downarrow$. Working within the quasi-particle approximation, we then developed idealized models to study the dynamics~\cite{Williams2004a} and thermodynamics~\cite{Williams2004b} of molecule formation. In this letter, we describe the salient features of this approach and apply our theory to the experiments of Hodby  \etal~\cite{Hodby2005a}.

In~\cite{Williams2004a}, we used our kinetic theory to derive rate equations for the atomic ($A$) and molecular ($M$) populations $N_j(t)=\int d \mathbf{r} \int  d \mathbf{p} f_j(\mathbf{p},\mathbf{r},t)/h^3$, where $j$ $=$ $A$ or $M$:
\begin{equation}
\label{NMdot}
\dot N_M(t) = \bar\gamma_f(t) N_A(t) -  \gamma_d(t) N_M(t), 
\end{equation}
and $\dot N_A(t) = -\dot N_M(t)$. The quantity $\gamma_d(t)$ is the molecular dissociation rate and $\bar\gamma_f(t)=\int d{\mathbf{r}}n_A({\mathbf{r}},t)\gamma_f({\mathbf{r}},t)/N_A(t)$ is the density-weighted average rate of molecule formation. The density is defined by $n_j({ \bf r},t)= \int d{\mathbf{p}}f_j({\mathbf{p}},{\mathbf{r}},t)/h^3$. The rates of molecule formation $\gamma_f({\bf r},t)$ and dissociation $\gamma_d(t)$ follow directly from the ``in" and ``out" terms of the atom-molecule collision integrals in the Boltzmann equations, which describe the relaxation toward chemical equilibrium due to resonant two-body collisions. Within the classical gas approximation (i.e. $f_i\ll 1$), they are
\begin{widetext}
\begin{eqnarray}
\label{gammad}
\gamma_d (t)&=& \frac{4 \pi^2 \, g^2}{h^7 n_M({\bf r},t)}  
\int d{\mathbf{p}}_1 \int d{\mathbf{p}}_2 \int d {\mathbf{p}}_3 \delta({\bf p}_1 + {\bf p}_2 - {\bf p}_3) 
\delta(\epsilon_A({\mathbf{p}}_1)+\epsilon_A({\mathbf{p}}_2)-\epsilon_M({\mathbf{p}}_3)) 
f_M({\bf p}_3) , \\
\label{gammaf}
\gamma_f ({\bf r},t) &=& \frac{4 \pi^2 \, g^2}{h^7 n_A({\bf r},t) } 
\int d{\mathbf{p}}_1 \int d{\mathbf{p}}_2 \int d {\mathbf{p}}_3 \delta({\bf p}_1 + {\bf p}_2 - {\bf p}_3) 
\delta(\epsilon_A({\mathbf{p}}_1)+\epsilon_A({\mathbf{p}}_2)-\epsilon_M({\mathbf{p}}_3)) 
 f_A({\bf p_1})f_A({\bf p}_2) ,  
\end{eqnarray}
\end{widetext}
where the explicit position and time dependence has been suppressed in the integrals. Neglecting self-energy effects, the atom and molecule energies are $\epsilon_A({\mathbf{p}},{\mathbf{r}})\equiv p^2/2m + U_{\rm{ext}}({\mathbf{r}})$ and $\epsilon_M({\mathbf{p}},{\mathbf{r}},t)\equiv p^2/4m + 2U_{\rm{ext}}({\mathbf{r}})+\epsilon_{\rm{res}}[B(t)]$, respectively, where $U_{\rm{ext}}({\mathbf{r}})$ is the external trapping potential and $\epsilon_{\rm{res}}[B(t)]$ is the renormalized energy of the resonant closed-channel state. This energy corresponds to the peak position of the molecule spectral function with the above-threshold limiting forms $\epsilon_{\rm{res}}(B) =\Delta \mu (B - B_0)$ far from resonance $|B-B_0|\gg |\Delta B|$ and $\epsilon_{\rm{res}}(B)\approx \hbar^2/m a^2(B)$ close to the resonance $|B-B_0|\ll |\Delta B|$~\cite{Nygaard2005a}. Here $B_0$ is the resonance position, $\Delta B$ is the resonance width, and $\Delta \mu$ is the magnetic moment difference between the open and closed channels. The scattering length near a Feshbach resonance is $a(B) = a_{\rm{bg}}[1-\Delta B/(B - B_0)]$, where $a_{\rm{bg}}$ is the background scattering length. Neglecting the energy dependence of interactions here, the Feshbach coupling strength $g$ is related to physical parameters according to $g^2=4\pi\hbar^2 a_{\rm{bg}} \Delta \mu \Delta B/m$. 

The rates in (\ref{gammad}) and (\ref{gammaf}) can be calculated explicitly when the system is in equilibrium $f_j({\mathbf{p}},{\mathbf{r}}) = \exp\{-[\epsilon_j({\mathbf{p}},{\mathbf{r}})-\mu_j]/k_{\rm{B}} T\}$, with $\mu_M = 2\mu_A \equiv 2\mu$. The molecular dissociation rate then simplifies to the position-independent form
\begin{equation}
\label{gammad2}
\gamma_d = \frac{2 m^{1/2}}{\hbar^2} a_{\rm{bg}} \Delta\mu \Delta B \sqrt{\epsilon_{\rm{res}}} \,\, \theta(\epsilon_{\rm{res}}) .
\end{equation}
Here, the unit step function $\theta(\epsilon_{\rm{res}})$ follows from energy and momentum conservation and signifies that below threshold $\epsilon_{\rm{res}}<0$, the pairs become stable and no longer can decay. We note that (\ref{gammad2}) is consistent with the decay rate derived in Ref.~\cite{Mukaiyama2004a}. The formation rate reduces to
\begin{equation}
\label{gammaf2}
\gamma_f ({\bf r})= 2^{3/2} \rho({\bf r}) e^{-\epsilon_{\rm{res}}/k_{\rm{B}}T} \gamma_d ,
\end{equation}
where $\rho({\bf r})\equiv \lambda_{\rm{th}}^3 n_A({\bf r})$ is the phase space density of the atomic component and $\lambda_{\rm{th}} = h/\sqrt{2\pi m k_{\rm{B}}T}$ is the thermal deBroglie wavelength.  The density-weighted average is $\bar\gamma_f = \gamma_f(0)/2^{3/2}$. The formation rate vanishes as $\epsilon_{\rm{res}}$ is tuned below threshold, in the same way that $\gamma_d$ does. This causes the saturation of molecule production during a Feshbach sweep: once $\epsilon_{\rm{res}}$ is tuned below threshold, no more molecules are formed via two-body collisions. 

Despite its simple intuitive form, Eq.~(\ref{NMdot}) cannot be solved without knowing the $f_j({\mathbf{p}},{\mathbf{r}},t)$ of Eq.~(\ref{gammad}) and Eq.~(\ref{gammaf}). Nonetheless, we can extract from Eq.~(\ref{NMdot}) an approximate form for the characteristic rate constant $\alpha$ that demarcates the adiabatic regime. At the very onset of molecule formation when $\chi\ll1$, the solution of Eq.~(\ref{NMdot}) can be approximated by $N_M(t_{\rm{ramp}})\approx (N_{\rm{tot}}/2)\{1 - \exp[-\int_{0}^{t_{\rm{ramp}}}\bar\gamma_f(\tau)d\tau] \}$. Making a change of variables from $\tau$ to $\epsilon_{\rm{res}}$ in the integral, and taking the large detuning form $\dot \epsilon_{\rm{res}} = -|\Delta \mu| |\dot B|$, we obtain $N_M(\dot B) \propto [1 - \exp (-\alpha/|\dot B|)]$, where the rate constant is $\alpha = \int_0^\infty d\epsilon_{\rm{res}} \bar\gamma_f(\epsilon_{\rm{res}})/|\Delta\mu|$. Evaluating the integral using the equilibrium form of $\bar\gamma_f$ gives~\cite{footnote2}
\begin{equation}
\label{alpha}
\alpha = 2^{3/2} \, \pi^2  \frac{\hbar a_{\rm{bg}}}{m}\frac{\Delta\mu}{|\Delta \mu|} \Delta B \, n_{A0},
\end{equation}
where $n_{A0}\equiv n_A({\bf r}=0)$ is the peak atomic density. This result (\ref{alpha}) resembles the Landau-Zener rate constant derived for a zero temperature Bose gas in~\cite{Goral2004a}. The experimental data in \cite{Hodby2005a} seem to be consistent with this linear dependence on density. 

In the adiabatic regime $\alpha/|\dot B| \gg 1$, the system follows an isentropic path as $\epsilon_{\rm{res}}$ is ramped downward~\cite{Williams2004a}.  When the resonance energy $\epsilon_{\rm{res}}[B(t)]$ is ramped to negative values, the two-body rates of molecule formation $\gamma_f$ and dissociation $\gamma_d$ vanish, at which point the process of molecule formation must proceed via three-body collisional relaxation, which may play a role for very slow ramps~\cite{Cubizolles2003a}.  When the timescale for three-body relaxation is much longer than $t_{\rm{ramp}}$, it can be neglected~\cite{Petrov2003a,Chin2003b}. Thus, molecule formation effectively halts when $\epsilon_{\rm{res}}$ crosses threshold at $B=B_0$, giving rise to the observed saturation of molecule production.

\begin{figure}[t]
\begin{center}
\includegraphics[scale=0.43]{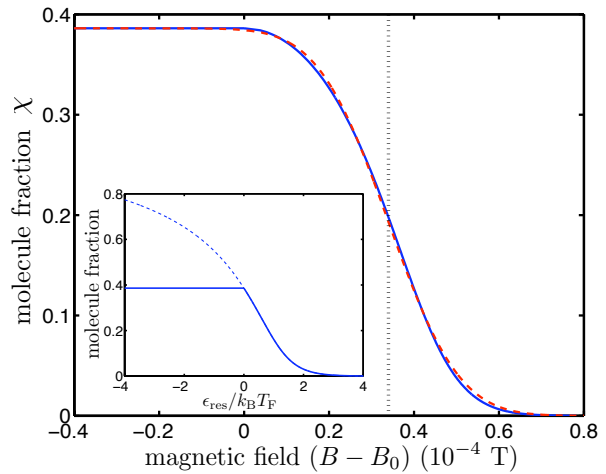}
\caption{Molecular formation curve for the case of ${}^{40}$K with $T_i/T_{\rm{F}}=0.5$. The solid blue line in both graphs corresponds to the constant entropy equilibrium solution. The dashed red line $\chi_{\rm{erf}}(B)$ with $\delta B = 0.19 \times 10^{-4}$ T and $B_{\rm{cen}}-B_0=0.34 \times 10^{-4}$ T. The dashed blue line in the inset shows the case where the system remains in chemical equilibrium for below threshold.}
\end{center}
\end{figure}

In~\cite{Williams2004b} we calculated constant entropy contours for a quantum ideal gas mixture of fermionic atoms and bosonic molecules without accounting for this saturation effect. It is straightforward to incorporate this effect into our earlier thermodynamic calculation. For positive detunings where $\epsilon_{\rm{res}}>0$, the results are unchanged: the molecular fraction $\chi(\epsilon_{\rm{res}})$ is determined according to the conservation of entropy and total atom number with the constraint that the atomic and molecular components are in thermal $T_M=T_A\equiv T$ and chemical $\mu_M = 2\mu_A \equiv 2\mu$ equilibrium. For negative detunings, the molecular fraction is held fixed $\chi(\epsilon_{\rm{res}}<0)=\chi(0)\equiv\chi_0$, where $\chi_0$ is the maximum production efficiency. The details of this type of calculation are given in~\cite{Williams2004b}.

In Fig. 2 we plot the molecular fraction along an isentropic path (solid blue lines) for the
specific case of an initial temperature $T_i/T_{\rm{F}}=0.5$, taking values for ${}^{40}$K: $B_0=202 \times10^{-4}$ T, $\Delta B = 7.4 \times10^{-4}$ T, and $a_{\rm{bg}}=174 a_0$~\cite{Greiner2003a}. The inset shows $\chi(\epsilon_{\rm{res}})$ versus $\epsilon_{\rm{res}}$ compared to the case where the gas remains in chemical equilibrium for negative detunings (dashed blue line). To plot $\chi$ versus magnetic field, the small detuning limit was used for $\epsilon_{\rm{res}}(B)$. The resulting molecular formation curve $\chi(B)$ is fit remarkably well by the error function $\chi_{\rm{erf}}(B)$ (dashed red line). The fitted width and centroid for four different initial temperatures $T/T_{\rm{F}}=\{0.25,0.5,0.75,1.0 \}$ are $\delta B = \{ 0.14, 0.19, 0.21, 0.24\}\times 10^{-4}$ T and $B_{\rm{cen}}-B_0=\{0.37, 0.34, 0.35, 0.36\}\times 10^{-4}$ T. These widths $\delta B$ are consistent with the value $0.2\times 10^{-4}$ T measured in experiments~\cite{Regal2003a,Regal2004a}, which is much smaller than the Feshbach resonance itself ($\Delta B = 7.4 \times10^{-4}$ T). A crucial point arising from our analysis is that the centroid $B_{\rm{cen}}$ of the error function does not coincide with the resonance position $B_0$. The centroid occurs roughly where $\epsilon_{\rm{res}}(B)\approx k_BT_{\rm{F}}$, so that
$B_{\rm{cen}}-B_0 \approx \Delta B a_{\rm{bg}} \sqrt{m k_{\rm{B}} T_{\rm{F}}}/\hbar$. 

Our saturation mechanism is quite general and occurs in both Bose and Fermi gases; examples of both systems have been studied by Hodby {\it{et al.}}~\cite{Hodby2005a}. Within the classical gas approximation it is straightforward to relate $\chi_0$ to the initial peak atomic phase space density $\rho_{0i}$. The maximum molecular fraction is given by $\chi_0 \equiv 2 N_M(T_f,\rho_{0f},\epsilon_{\rm{res}}=0)/N_{\rm{tot}}$, where $T_f$ and $\rho_{0f}$ are the temperature and peak atomic phase space density at $\epsilon_{\rm{res}}=0$. The total atomic population is $N_{\rm{tot}}=\eta N_A + 2N_M$, where for generality we allow for the number of atomic spin components to be $\eta\in\{1,2\}$.  The atomic and molecular populations can be written as $N_A(T,\rho_0) = (k_{\rm{B}}T/\hbar \bar\omega)^3 \rho_0$ and $N_M(T,\rho_0,\epsilon_{\rm{res}}) = (k_{\rm{B}}T/\hbar \bar\omega)^3 \rho_0^2 \exp(-\epsilon_{\rm{res}}/k_{\rm{B}}T)$, where the mean trapping frequency is $\bar\omega=(\omega_x\omega_y\omega_z)^{1/3}$. Using these expressions, the molecular fraction takes the form $\chi_0 = \rho_{0f}/(\eta/2+\rho_{0f})$. We can relate $\rho_{0f}$ to $\rho_{0i}$ using entropy conservation. The atomic and molecular entropies can be written as $S_A(T,\rho_0)=k_{\rm{B}}N_A(4 - \ln \rho_{0})$ and $S_M(T,\rho_0,\epsilon_{\rm{res}})=k_{\rm{B}}N_M(4 + \epsilon_{\rm{res}}/k_{\rm{B}}T- 2\ln \rho_{0})$.  Substituting these into the entropy conservation equation $\eta S_A(T_i,\rho_{0i})=\eta S_A(T_f,\rho_{0f})+S_M(T_f,\rho_{0f},\epsilon_{\rm{res}}=0)$ yields the relation $\ln\rho_{A0i}=\ln\rho_{A0f}+2\chi_0$.  From these observations, the following key result can be derived
\begin{equation}
\label{conversion}
2\chi_0 + \ln \left (\frac{\eta}{2}\, \frac{\chi_0}{1-\chi_0} \right ) = \ln \rho_{0i} .
\end{equation}

The solution $\chi_0$ of this transcendental equation is a universal function of the 
the initial peak atomic phase space density $\rho_{0i}$ and the number of spin components $\eta$.  A similar derivation for an ideal quantum gas results in a much more complicated expression, in which $\chi_0$ also depends on the quantum statistics of the different components. In Fig. 3 we compare our theoretical prediction of $\chi_0$, both with (solid blue) and without (dashed green) quantum statistical effects, to experimental data and find qualitative agreement with the JILA data~\cite{Regal2003a,Hodby2005a} but not with the Rice data point (red diamond)~\cite{Strecker2003a}.

\begin{figure}[t]
\begin{center}
\includegraphics[scale=0.43]{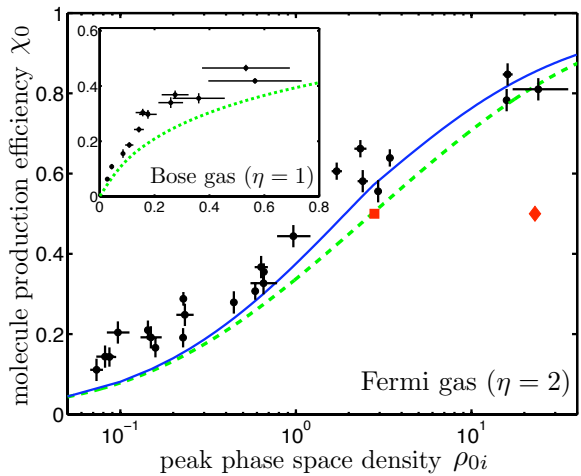}
\caption{Comparison between theory and experiment. The solid blue curve is for a two-component Fermi gas including quantum statistical effects. The dashed green line shows the classical result, Eq.~(\ref{conversion}), for $\eta=2$. The solid black circles are experimental data for ${}^{40}$K from Hodby {\it{et al.}}~\cite{Hodby2005a}. The two earliest experiments are those of JILA (red square)~\cite{Regal2003a} and Rice (red diamond)~\cite{Strecker2003a}. The inset compares the $\eta=1$ case of Eq.~(\ref{conversion}) (dashed green line) to data for ${}^{85}$Rb taken in the classical regime of low phase space density~\cite{Hodby2005a}.}
\end{center}
\end{figure}

In summary, we have treated Feshbach molecule formation in a dilute gas during a magnetic field ramp as a relaxation process described by kinetic theory~\cite{Williams2004a,Nygaard2005b} and thermodynamics~\cite{Williams2004b}. Such a treatment is appropriate to the conditions of present experiments~\cite{Regal2003a,Strecker2003a,Hodby2005a}. Our key findings are: (i) the rate factor $\alpha$ is proportional to density, (ii) the width $\delta B$ and centroid $B_{\rm{cen}}$ of the formation curve are related to the temperature $T$ of the gas, and (iii) the molecule production efficiency $\chi_0$ is a universal function of the initial peak atomic phase space density $\rho_{0i}$, which in the classical regime is approximated by the ideal gas result Eq.~(\ref{conversion}). Our predictions are consistent with most of the existing experimental data.

We thank E. Hodby and C. Regal for providing the experimental data used in Fig. 3. We appreciate useful discussions with  T. Nikuni, P. Julienne and E. Tiesinga.

\end{document}